\documentclass{PoS}

\title{Neutrinos, Fission Cycling, and the $r$-process}

\ShortTitle{Neutrinos, Fission Cycling, and the $r$-process}

\author{\speaker{J. Beun}\\
        Department of Physics, North Carolina State University, Raleigh, NC 27695-8202\\
        E-mail: \email{jbbeun@unity.ncsu.edu}}

 \author{G. C. McLaughlin\\
 Department of Physics, North Carolina State University, Raleigh, NC 27695-8202}
\author{R. Surman       \\
  Department of Physics, Union College, Schenectady, NY 12308}      
\author{W. R. Hix       \\
Physics Division, Oak Ridge National Laboratory, Oak Ridge, TN, 37831-6374}

\abstract{It has long been suggested that fission cycling may play an important
role in the $r$-process.  Fission cycling can only occur in a very neutron
rich environment.  In traditional calculations of the neutrino driven
wind of the core-collapse supernova, the environment is not
sufficiently  neutron rich to produce the $r$-process elements. However,
we show that with a reduction of the electron neutrino flux coming from
the supernova, fission cycling does occur and furthermore it produces
an abundance pattern which is consistent with observed $r$-process
abundance pattern in halo stars.  Such a reduction can be caused by
active-sterile neutrino oscillations or other new physics.
}

\FullConference{International Symposium on Nuclear Astrophysics --- Nuclei in the Cosmos --- IX\\
		 June 25-30 2006\\
		 CERN, Geneva, Switzerland}

\begin{document}

\section{Introduction}

The search for the $r$-process site has endured for some time now,
yielding several promising locations without a clear answer \cite{1994ApJ...433..229W,1992ApJ...399..656M,2003MNRAS.345.1077R,2004ApJ...603..611S}.
The neutrino driven wind of the core-collapse supernova remains promising
and is attractive from the point of view of timescale arguments \cite{2001ApJ...554..578W}; 
however, there still remain some unresolved difficulties, for example, those stemming from entropy conditions \cite{1994A&A...286..857T}
and the $\alpha$ effect \cite{1997ApJ...482..951H,1998PhRvC..58.3696M}.
In the core-collapse supernova environment, several physical modifications have been proposed to achieve an $r$-process,
 such as a fast outflow wind 
\cite{1997ApJ...482..951H}, and active-sterile neutrino oscillations \cite{1999PhRvC..59.2873M,Beun:2006ka}.
A reduction in the electron neutrino,$\nu_e$, flux would be effective for producing a suitable environment 
for the nucleosynthesis of $r$-process elements.  In this contribution, we investigate the impact of such a reduction
without tying it to a particular mechanism.  Although the outflow would be diminished by a lower neutrino luminosity, 
effects such as multi-dimensional effects might compensate for this.  
Additionally, active-sterile neutrino oscillations occur after the neutrino energy deposition has occurred,
 leaving the outflow unmodified.

\section{Calculation and Comparison with Data}

\begin{figure}[h]
\centerline{\includegraphics[height=2.5in]{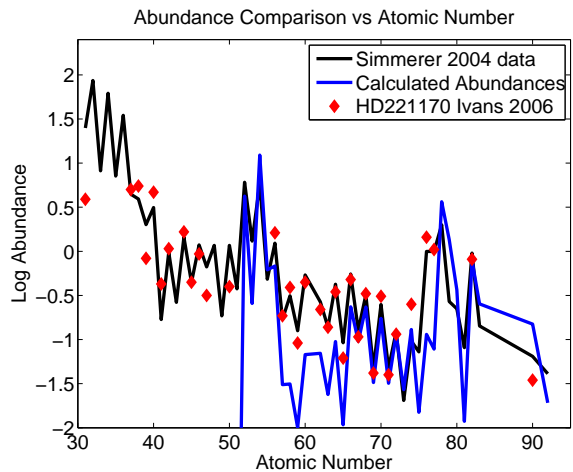}}
\caption{Comparison of the abundance pattern produced by our full network calculation in a post-core bounce supernova environment environment 
with the solar system \protect \cite{Simmerer:2004jq} and halo star data \protect \cite{Ivans:2006sj}. 
The neutrino and wind parameters were $L_\nu = 0.8 \times 10^{50} \, {\rm erg} \, {\rm s}^{-1}$
 and $L_{\bar{\nu}} = 4 \times 10^{51} \, {\rm erg} \, {\rm s}^{-1}$, $s/k = 100$ with an outflow timescale of $\tau = 0.3 s$.}
\label{y}
\end{figure}

Observations of halo star abundances have yielded an $r$-process pattern remarkably similar to that of the solar system 
\cite{2003Sci...299...70S},
most strongly correlated in the region between the second and third $r$-process peaks.  In a nucleosynthesis calculation in the neutrino driven wind of the supernova with the 
$\nu_e$ flux from the proto-neutron star reduced by a factor of $\sim 10$, we find an $r$-process abundance pattern 
in rough correspondence to the second and third peak regime of the solar system $r$-process, as shown in Fig. \ref{y}.
Our agreement range coincides with that of the halo star data; strengthening the connection with
supernovae, which would occur early in the evolution of the universe.

\section{Fission Cycling}

\begin{figure}[h]
\centerline{\includegraphics[height=2.5in]{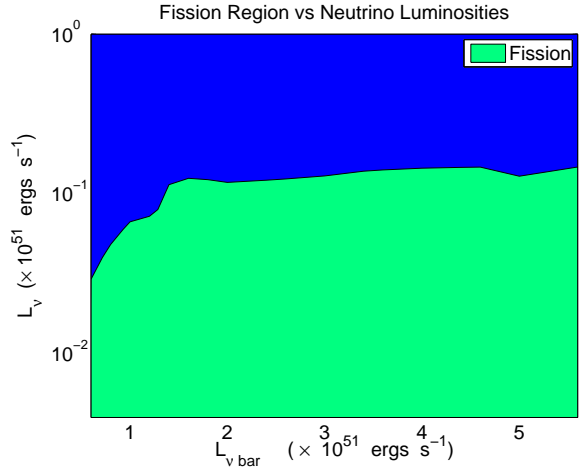}}
\caption{The onset of fission as a function of supernova neutrino luminosities.
The calculations shown in this figure are of the neutrino driven wind environment of core collapse supernovae.  The neutrino average energies are
$\langle E_{\nu_e} \rangle = 11 \; {\rm MeV}$ and $\langle E_{\bar{\nu}_e} \rangle = 16 \; {\rm MeV}$.  }
\label{fis}
\end{figure}

A mechanism within the $r$-process responsible for generating solely the nuclei above and between the second and third $r$-process peaks is fission cycling.  
While the role of fission has been previously studied, \emph{e.g.}
\cite{1991PhR...208..267C}, 
the exact details about which nuclei undergo fission and their daughter products are unknown.
Generally, heavy fissionable nuclei are expected to produce daughters with atomic weights greater than the first $r$-process peak
(A $\approx$ 80 \& Z $\approx$ 35).
During fission cycling neutron capture not only builds up nuclei to such large atomic number that fission occurs, there are still plenty of neutrons to capture on the daughter products of the fissioned nucleus. 
The daughters effectively become new $r$-process seed nuclei.  It is this build up of the daughters to high atomic mass number, followed by fission creates the cycle.
In Fig \ref{fis}. we show that 
the onset of fission is established for electron neutrino luminosities of $L_{\nu_e} < 10^{50} \, {\rm erg} \, {\rm s}^{-1}$ 
and for electron anti-neutrino luminosities of $L_{{\bar{\nu}}_e} > 10^{51} {\rm erg} \, {\rm s}^{-1}$. 

\section{Steady $\beta$ Flow}

\begin{figure}[h]
\centerline{\includegraphics[height=2.5in]{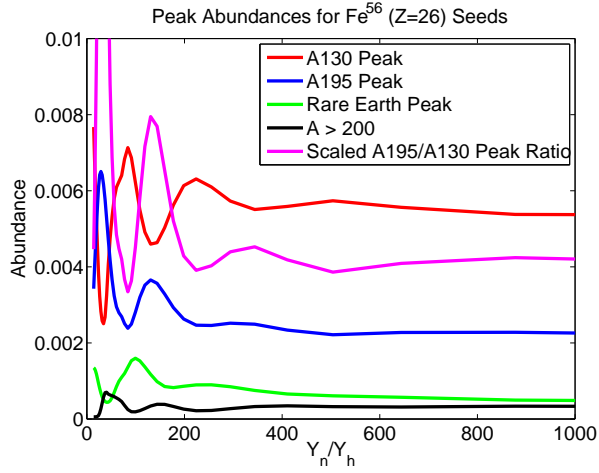}}
\caption{The $r$-process peak heights as a function of neutron-to-seed ratio, for an exploratory calculation that begins the r-process 
with $^{56}{\rm Fe}$ seeds.  Above a neutron to seed ratio of $Y_n/Y_h \approx 400$,
 fission cycling occurs and final peak heights are independent of the neutron to seed ratio.}
\label{ynpeak}
\end{figure}

We evaluate the robustness of an $r$-process pattern which comes about due to fission cycling
by considering a schematic model of $^{56}{\rm Fe}$ seeds and a neutron excess.
Peak heights are calculated and plotted against neutron to seed ratio, $Y_n/Y_h$ in Fig. \ref{ynpeak}.  As the neutron abundance is increased to a critical point, $Y_n/Y_h \sim 400$, the second and third $r$-process peaks stabilize to constant values.

This stability comes about from fission cycling,
 which allows the material to establish steady $\beta$ flow.
Steady $\beta$ flow, see \emph{e.g.} \cite{1991PhR...208..267C} for a review,
comes about when
electromagnetic and strong interactions happen on a much faster timescale than beta decay. 
Neutron-capture and photo-disintegration rates equilibrate marking (n,$\gamma$) $\rightleftarrows$ ($\gamma$,n) equilibrium.
Once (n,$\gamma$) $\rightleftarrows$ ($\gamma$,n) equilibrium occurs, the change in abundance, $Y(Z,A)$, can be described in terms of
the beta decay rates, $\lambda_{_\beta}$, between two adjacent chains in $Z$,

\begin{equation}\label{bflow1}
\dot{Y}(Z) =  \sum_{A}^{} Y(Z-1,A)\lambda_{\beta}(Z-1,A) -  \sum_{A}^{} Y(Z)\lambda_{\beta}(Z,A).
\end{equation}

From this, the condition for steady $\beta$ flow is 

\begin{equation}\label{bflow2}
\sum_{A}^{} Y(Z,A)\lambda_{\beta}(Z,A) = const.
\end{equation}

\begin{figure}[h]
\centerline{\includegraphics[height=2.5in]{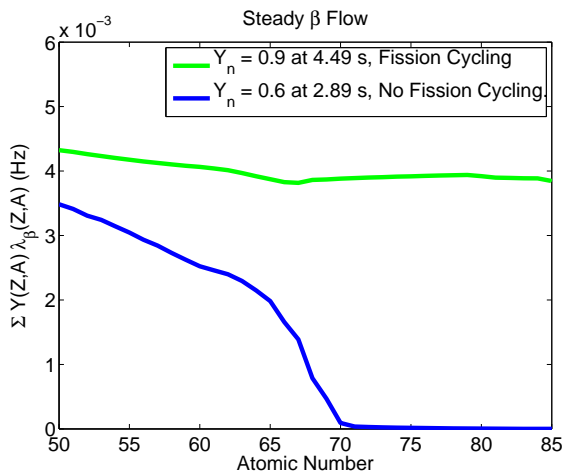}}
\caption{
Plot of abundance times the beta decay rate, $\sum_{A}^{} Y(Z,A)\lambda_{\beta}(Z,A)$, 
for the same type of calculation shown in Fig. \protect \ref{ynpeak} for two different neutron excesses.  
The nearly constant value at high neutron excess means that the constant peak heights
in Fig. \protect \ref{ynpeak} can be attributed to steady $\beta$ flow.}
\label{path}
\end{figure}

Freeze-out from steady $\beta$ flow realizes a consistent $r$-process abundance pattern.
The achievement of steady $\beta$ flow is shown in Fig. \ref{path}, 
designated by the straight line of the more neutron-rich case in comparison with the less neutron-rich case.
Also important is that steady $\beta$ flow occurs over the region between the second and third $r$-process peaks;
further demonstrating connection with the halo star $r$-process data.

\section{Conclusion}
Fission cycling and steady $\beta$ flow are promising mechanisms for establishing 
a robust second and third $r$-process peak as seen in halo stars.  
In the neutrino driven wind of the supernova, this can come about due to a decrease in the predicted electron neutrino flux. 
 In neutrino driven wind models with $\tau = 0.3$ s and $s/k = 100$, we find fission cycling occurs for 
$L_{\nu_e} < 10^{50} \, {\rm erg} \, {\rm s}^{-1}$ and $L_{\overline{\nu}_e} > 10^{51} {\rm erg} \, {\rm s}^{-1}$. 
These luminosities can come about from active-sterile neutrino oscillations, but they may perhaps be generated from other new physics.


%
%
%
%


\def\ref@jnl#1{{#1}}

\def\aj{\ref@jnl{AJ}}                   
\def\araa{\ref@jnl{ARA\&A}}             
\def\apj{\ref@jnl{ApJ}}                 
\def\apjl{\ref@jnl{ApJ}}                
\def\apjs{\ref@jnl{ApJS}}               
\def\ao{\ref@jnl{Appl.~Opt.}}           
\def\apss{\ref@jnl{Ap\&SS}}             
\def\aap{\ref@jnl{A\&A}}                
\def\aapr{\ref@jnl{A\&A~Rev.}}          
\def\aaps{\ref@jnl{A\&AS}}              
\def\azh{\ref@jnl{AZh}}                 
\def\baas{\ref@jnl{BAAS}}               
\def\jrasc{\ref@jnl{JRASC}}             
\def\memras{\ref@jnl{MmRAS}}            
\def\mnras{\ref@jnl{MNRAS}}             
\def\pra{\ref@jnl{Phys.~Rev.~A}}        
\def\prb{\ref@jnl{Phys.~Rev.~B}}        
\def\prc{\ref@jnl{Phys.~Rev.~C}}        
\def\prd{\ref@jnl{Phys.~Rev.~D}}        
\def\pre{\ref@jnl{Phys.~Rev.~E}}        
\def\prl{\ref@jnl{Phys.~Rev.~Lett.}}    
\def\pasp{\ref@jnl{PASP}}               
\def\pasj{\ref@jnl{PASJ}}               
\def\qjras{\ref@jnl{QJRAS}}             
\def\skytel{\ref@jnl{S\&T}}             
\def\solphys{\ref@jnl{Sol.~Phys.}}      
\def\sovast{\ref@jnl{Soviet~Ast.}}      
\def\ssr{\ref@jnl{Space~Sci.~Rev.}}     
\def\zap{\ref@jnl{ZAp}}                 
\def\nat{\ref@jnl{Nature}}              
\def\iaucirc{\ref@jnl{IAU~Circ.}}       
\def\aplett{\ref@jnl{Astrophys.~Lett.}} 
\def\apspr{\ref@jnl{Astrophys.~Space~Phys.~Res.}}
\def\bain{\ref@jnl{Bull.~Astron.~Inst.~Netherlands}} 
\def\fcp{\ref@jnl{Fund.~Cosmic~Phys.}}  
\def\gca{\ref@jnl{Geochim.~Cosmochim.~Acta}}   
\def\grl{\ref@jnl{Geophys.~Res.~Lett.}} 
\def\jcp{\ref@jnl{J.~Chem.~Phys.}}      
\def\jgr{\ref@jnl{J.~Geophys.~Res.}}    
\def\jqsrt{\ref@jnl{J.~Quant.~Spec.~Radiat.~Transf.}}
\def\memsai{\ref@jnl{Mem.~Soc.~Astron.~Italiana}}
\def\nphysa{\ref@jnl{Nucl.~Phys.~A}}   
\def\physrep{\ref@jnl{Phys.~Rep.}}   
\def\physscr{\ref@jnl{Phys.~Scr}}   
\def\planss{\ref@jnl{Planet.~Space~Sci.}}   
\def\procspie{\ref@jnl{Proc.~SPIE}}   

\let\astap=\aap
\let\apjlett=\apjl
\let\apjsupp=\apjs
\let\applopt=\ao


\bibliographystyle{unsrt}
\bibliography{master}


\end{document}